\documentclass[conference]{IEEEtran}
\IEEEoverridecommandlockouts
\usepackage{cite}
\usepackage{amsmath,amssymb,amsfonts}
\usepackage{algorithmic}
\usepackage{graphicx}
\usepackage{textcomp}
\usepackage{xcolor}
\def\BibTeX{{\rm B\kern-.05em{\sc i\kern-.025em b}\kern-.08em
    T\kern-.1667em\lower.7ex\hbox{E}\kern-.125emX}}

\usepackage{cite}
\usepackage{subcaption}
\usepackage{optidef}
\usepackage[hyphens]{xurl} 
\usepackage{color}
\usepackage{comment}

\newcommand{\add}[1]{\textcolor{black}{#1}}

\begin{document}

\title{Experimental Realization\\ of Koopman-Model Predictive Control\\
for an AC-DC Converter}

\author{\IEEEauthorblockN{Shun Hirose}
\IEEEauthorblockA{\textit{Dept. of Electrical Engineering} \\
\textit{Kyoto University}\\
Kyoto, Japan \\
hirose.shun.65x@st.kyoto-u.ac.jp}
\and
\IEEEauthorblockN{Shiu Mochiyama}
\IEEEauthorblockA{\textit{Dept. of Electrical Engineering} \\
\textit{Kyoto University}\\
Kyoto, Japan \\
mochiyama.shiu.5c@kyoto-u.ac.jp}
\and
\IEEEauthorblockN{Yoshihiko Susuki}
\IEEEauthorblockA{\textit{Dept. of Electrical Engineering} \\
\textit{Kyoto University}\\
Kyoto, Japan \\
susuki.yoshihiko.5c@kyoto-u.ac.jp}
}

\maketitle

\begin{abstract}
This paper experimentally demonstrates the Koopman-Model Predictive Control (K-MPC) for a real AC-DC converter. 
The converter is typically modeled with a nonlinear time-variant plant. 
We introduce a new dynamical approach to lifting measurable dynamics from the plant and constructing a linear time-invariant model that is consistent with control objectives of the converter.
We show that the lifting approach, combined with the K-MPC controller, performs well across the full experimental system and outperforms existing control strategies in terms of both steady-state and transient responses.
\end{abstract}


\section{Introduction}\label{introduction}

Control of power converters has attracted considerable attention due to increasing demand for high-performance capabilities in emerging applications \cite{IEEE-Electrific-Mag,IEEE-Power-Electron-Mag}. 
The AC-DC power conversion, which we will study in this paper, is a basic function for power supply technologies such as battery management systems and motor drive systems. 
There are numerous developments and control practices for AC-DC conversion, including PI control \cite{Euzeli}, nonlinear control such as the port-Hamiltonian approach \cite{ijcta}, and Model Predictive Control (MPC) \cite{mpc-for-converter}. 
Beyond its conventional use in battery management and motor drive, the so-called More Electric Aircraft (MEA) has necessitated high control performance for AC-DC conversion as an emerging application, see, e.g., \cite{onboard,barzkar}. 
MEA aims to develop an autonomous microgrid operating in a dynamic, uncertain, and extreme flight environment and to ensure reliable supply and high-quality power in the grid, including numerous facilities with complex dynamics. 
For the emerging application, it has been necessary to develop a new control strategy for AC-DC conversion that addresses its nonlinear and uncertain nature as a challenge in control systems technology.

This paper applies the Koopman-MPC (K-MPC) \cite{KORDA2018149} to a real AC-DC converter. 
The K-MPC for a given nonlinear plant utilizes a \emph{linear} model for predicting the output of the nonlinear plant, which is guided by the Koopman operator framework \cite{koopman-introduction-springer}. 
The linear modeling, which we call the Koopman Operator-based Modeling (KOM), is conducted by increasing the dimension of the output space with a combination of user-defined functions, called \emph{observables}, and input/output time-series data. 
KOM and K-MPC are capable of handling the nonlinearity of the target's plant and have gained success in power conversion systems 
\cite{hanke2019,koopman-dcdcconverter,Debnath2024,HUO2025106225}. 

We present the first experimental demonstration of K-MPC for a real AC-DC converter. 
The novel contributions of this paper are twofold. 
The control objectives here are concerned with the amplitude and phase of the AC current, and the mean value of the DC voltage, using direct measurement of their instantaneous current and voltage. 
To achieve the objectives using the KOM and K-MPC, as the first contribution, we introduce the idea of harmonic-average from \cite{gssa-introduction}, called the Generalized State-Space Averaging (GSSA), to the KOM for the AC-DC converter. 
This is a dynamical approach to synthesizing the so-called dictionary of observables in the KOM or extended Dynamic Mode Decomposition (eDMD) \cite{edmd-introduction}. 
Our approach is clearly different from the eDMD framwork in which a static map is normally incorporated as the observable \cite{Brunton_Kutz_2022}. 
We will show that the dynamical approach enables the K-MPC to achieve the control objectives of the AC-DC converter. 
As the second contribution, we present experimental data on the K-MPC for a real single-phase AC-DC converter comprising a full-bridge boost rectifier, as shown in Figs.~\ref{cir:1} and \ref{fig:photo}. 
The KOM used for the K-MPC is data-driven; that is, it is based solely on measured data from the real converter. 
These are novel points by comparison with the existing references \cite{hanke2019,koopman-dcdcconverter,Debnath2024,HUO2025106225}. 
To the best of our knowledge, the fully experimental application of the K-MPC to a real converter is novel. 
We will experimentally demonstrate that the K-MPC combined with the dynamical approach in KOM outperforms existing techniques, including PI control. 

\section{Prototype AC-DC Converter}\label{ac-dc-converter}

\begin{figure}[t]
    \centering
    \includegraphics[width=1.0\linewidth]{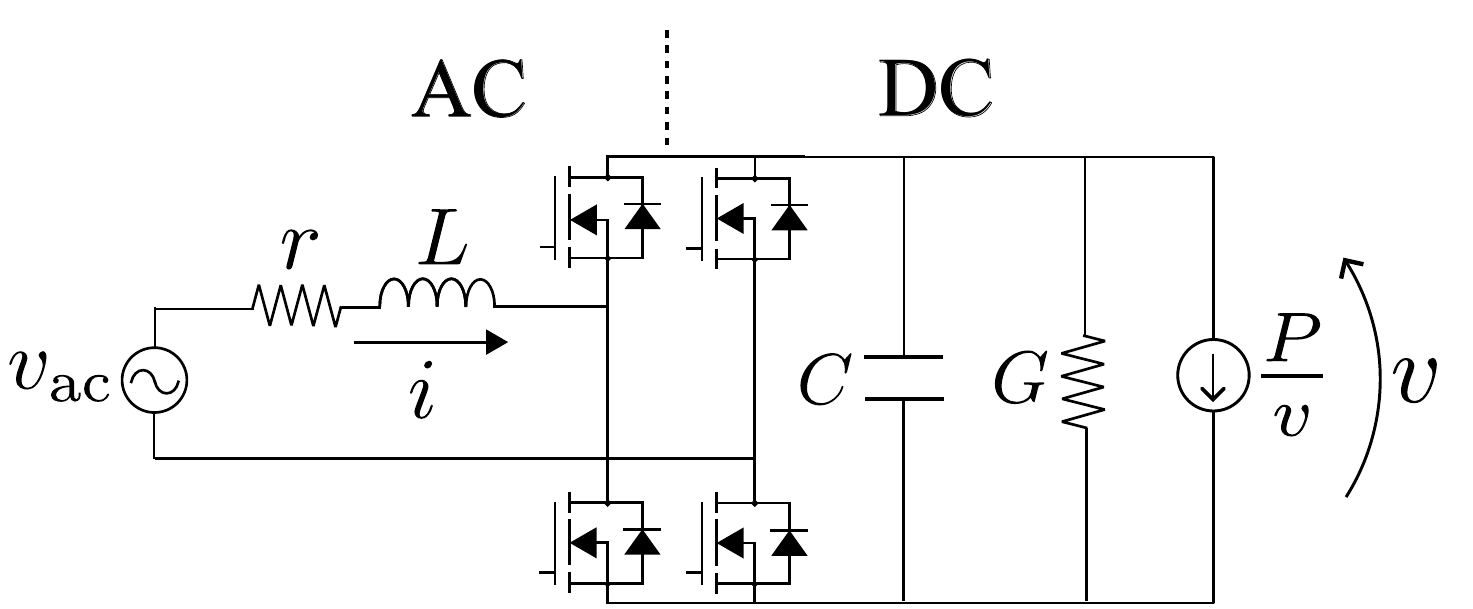}
    \caption{Basic circuit of an AC-DC converter. 
    The Constant Power Load (CPL) is represented by the symbol of current source. }
    \label{cir:1}
\end{figure}

\begin{figure}[t]
      \centering
      \includegraphics[width=1.0\linewidth]{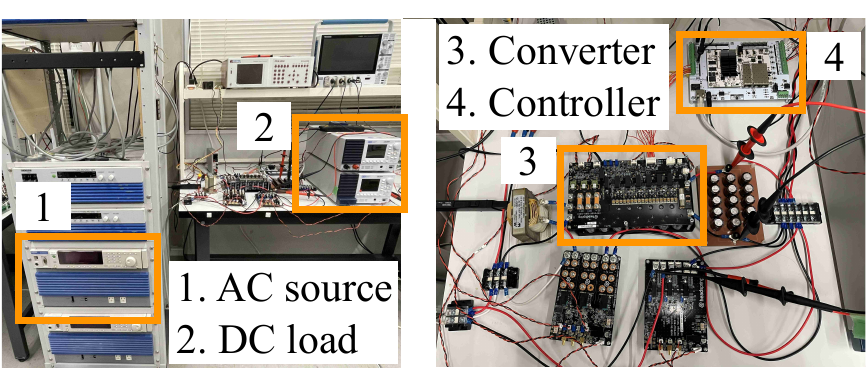}
      \caption{Photographs of the experimental system.}
      \label{fig:photo}
   \end{figure}

Figure~\ref{cir:1} shows a single-phase AC-DC converter containing a full-bridge boost rectifier. 
Photographs of the experimental system are presented in Fig.~\ref{fig:photo}. 
The four MOSFETs in the center of Fig.~\ref{cir:1} play a central role in converting the AC power into the DC power consumed by loads. 
The AC supply voltage $v_{\rm ac}(t)$ in continuous time $t$ is represented as $v_{\rm ac}(t) = E\sin(\omega t)$ with the amplitude $E=28\sqrt{2}\,\rm V$ and the angular frequency $\omega=2\pi\cdot50\,\rm rad/s$ (period $T=2\pi/\omega=20\,\rm ms$). 
The DC loads consist of a resistive load with $G=0.01\,\rm S$ and a Constant Power Load (CPL) with $P=25\,\rm W$ \add{, which is assumed to be unknown to address an uncertain environment for MEA as mentioned in Sec.~\ref{introduction}}. 
The current of the CPL is represented as $i_{\rm CPL}(t) =P/v(t)$, where $v(t)$ is the load voltage. 
The inductance of the AC part is $L=1\,\rm mH$, and the capacitor of the DC part is $C=4560\,\mu\rm F$. 
The resistance $r$ accounts for all resistive losses (inductor, source, and power devices). 
From measurement, we assume $r=80\,\rm m\Omega$ that is the sum of $60\,\rm m\Omega$ (MOSFETs' resistance) and $20\,\rm m\Omega$ (line resistance). 
The AC current $i(t)$ and the DC voltage $v(t)$ are measurable using two sensors 
connected to the converter. 

The converter's bridge, consisting of four SiC MOSFETs, is driven by Pulse-Width Modulation (PWM). 
The PWM signal $s(t)\in\{-1,1\}$ is generated in an FPGA \add{of the controller (B-Board from Imperix Ltd.)} used for implementing the K-MPC controller. 
The switching frequency (that is, the frequency of the carrier signal) is set at $20\,\rm kHz$, and the dead-time at $400\,\rm ns$. 
When $s(t)$ takes the value $1$ (or $-1$), the upper left (or right) and lower right (or left) MOSFETs turn on. 
The $s(t)$ is determined with the duty ratio $\mu(t)\in[-1,1]$ that is the control signal and generated in a CPU \add{of B-Board}. 

The control objectives for the AC-DC converter are listed as follows:
\begin{enumerate}
    \item The mean value of the DC voltage $v(t)$ is equal to the desired value $V_{\rm d}=48\,\rm V$.
    \item The power factor of the converter is equal to one; precisely, $i(t)$ becomes $I_{\rm d}\sin\omega t$ with constant amplitude $I_{\rm d}$. 
\end{enumerate}
Here, $I_{\rm d}$ is the unknown parameter. 
If the two control objectives are achieved, then the following power-balance equation holds (see \cite{ijcta} for derivation):
\begin{equation}
    \label{eq:power-balance}
    \frac{1}{2}EI_{\rm d} = \frac{1}{2}rI_{\rm d}^{2} + GV_{\rm d}^{2} + P.
\end{equation}
It is shown in \cite{ijcta} that the feasible value of $I_{\rm d}$, as the smaller solution of \eqref{eq:power-balance}, becomes $I_{\rm d}=2.44\,\rm A$. 
Furthermore, as described in Sec.~\ref{introduction} about the reliable power supply in the AC side, it is required to set the AC current limit motivated by the protection of semiconductor devices 
\cite{Peyghami2020}. 
From the fact that the overloading current of silicon devices is $1.5$ to $2$ times the rated current \cite{Peyghami2020}, we limit the AC current $i(t)$ to $1.5I_{\rm d}\sim4\,\rm A$, that is, 
\begin{equation}
|i(t)| \leq  1.5I_{\rm d}\sim 4\,\rm A.
\label{eq:limit}
\end{equation}

\section{Controller Design}\label{controller-design}

\subsection{Overview}

We apply the K-MPC (Koopman-Model Predictive Control) \cite{KORDA2018149} to the AC-DC converter. 
An overview of the application is presented here. 
The basic idea is to estimate a linear model using the KOM and to recursively solve the optimization problem subject to predictions generated by the linear model. 
The estimation is based on lifting measurable dynamics of the plant, denoted as $\boldsymbol{y}[k]$ ($k=0,1,2,\ldots$), where $k$ is the discrete time corresponding to the control onset at every period $T$, onto a high-dimensional space via a user-defined map. 
The $\boldsymbol{y}[k]$ in this paper are the measured data on the AC current $i(t)$ and the DC voltage $v(t)$. 
We here represent the lifted dynamics by $\boldsymbol{z}[k]$. 
Then, the linear model to be estimated is given by 
\begin{equation}
    \label{eq:linear-representation}
    \boldsymbol{z}[k+1] = \boldsymbol{A}\boldsymbol{z}[k] + \boldsymbol{B}\boldsymbol{u}[k],
\end{equation}
where $\boldsymbol{A}$ and $\boldsymbol{B}$ are finite-dimensional constant matrices, and $\boldsymbol{u}$ is the control input to be determined in the K-MPC controller. 
The estimation of $\boldsymbol{A}$ and $\boldsymbol{B}$ is based on the datasets $\{\boldsymbol{y}[1],\boldsymbol{y}[2],\ldots\}$ and $\{\boldsymbol{u}[1],\boldsymbol{u}[2],\ldots\}$ which are obtained from the collection of the measurable dynamics $\boldsymbol{y}$ and exogeneous inputs $\boldsymbol{u}$. 
\add{The estimation algorithm follows \cite{KORDA2018149}.}
The optimization problem in the K-MPC is formulated by incorporating the linear model \eqref{eq:linear-representation} as an equality constraint into the standard linear MPC: see \cite{KORDA2018149}. 

\subsection{Synthesis of the Lifted Dynamics}
\label{synthesis-of-lifted-dynamics}

The key to the KOM is to introduce the map for synthesizing the lifted dynamics $\boldsymbol{z}$. 
Normally, in the K-MPC, the so-called \emph{static} map from $\boldsymbol{y}$ to $\boldsymbol{z}$ is used, known as the observable or dictionary functions \cite{Brunton_Kutz_2022}. 
In this paper, we contend that the static map is unsuitable for achieving the control objectives with the K-MPC.  
The control objective 1) is to regulate the mean value of the DC voltage $v(t)$, and the mean value is never represented by the static map with no memory of $v[k]$. 
In addition, the control objective 2) is to track the time-varying AC current $i(t)$, and the tracking problem is never formulated in the time-invariant manner of \eqref{eq:linear-representation}. 
Note that these limitations are mentioned in \cite{ijcta} when the port-Hamiltonian approach is applied. 

To overcome these limitations, we propose using the so-called dynamic map to synthesize the lifted dynamics $\boldsymbol{z}$. 
Specifically, we incorporate the Generalized State-Space Averaging (GSSA) \cite{gssa-introduction}, which was originally used in \cite{ijcta} to apply the passivity-based control to an AC-DC converter. 
The GSSA is introduced here in discrete time. 
Regarding the sampling strategy, for a scalar measurement $y(t)$ ($i(t)$ or $v(t)$ in this paper), $y(j\Delta t)$ with $\Delta t=T/N$ ($N$ is an integer) represents the sampled data on $y(t)$ labeled by integer $j$, under uniform sampling of period $\Delta t$. 
This $\Delta t$ is smaller than the period $T$ of the AC voltage. 
It should be stated that $T$ corresponds to the control period, and the control onset in discrete time is labeled by integer $k$. 
Then, the index-$h$ GSSA at discrete time $k$ is defined as
\begin{equation}
    \label{eq:computeGSSA}
    \langle y \rangle_{h}[k] := \frac{1}{N} \sum_{j=(k-1)N}^{kN} y(j\varDelta t)\exp({-\mathrm{j}\omega hj\varDelta t}),
\end{equation}
where $\mathrm{j}=\sqrt{-1}$ stands for the imaginary unit.  
The $\langle y \rangle_h[k]$ at discrete time $k$ is the moving harmonic-average of the time series $\{y((k-1)N\varDelta t),\ldots,y(kN\varDelta t)\}$ updated over $T$. 
The $\langle i \rangle_{1}$ is related to the phasor of the AC current $i(t)$, and $\langle v \rangle_{0}$ is related to the mean value of $v(t)$. 
In addition, the reference signals of $i(t)$ and $v(t)$ are encoded with the GSSA as $\langle I_{\rm d}\sin(\omega t)\rangle_{1}=-\mathrm{j}I_{\rm d}/2$ and $\langle V_{\rm d}\rangle_{0}=V_{\rm d}$, which will be used for the K-MPC controller. 
In this paper, we synthesize the lifted dynamics as
\begin{equation}
    \label{eq:observables}
    \boldsymbol{z}[k]=[\Im\langle i \rangle_{1}[k],\Re\langle i \rangle_{1}[k],\langle v \rangle_{0}[k],\langle v^{-1} \rangle_{0}[k]]^\top,
\end{equation}
where $\Im(\cdot)$ and $\Re(\cdot)$ stands for the imaginary and real part of $(\cdot)$, respectively. 
The symbol $\top$ stands for the transpose operation of vectors. 
The two elements $\langle i \rangle_{1}$ and $\langle v \rangle_{0}$ are the outputs to be controlled, and the last element $\langle v^{-1}\rangle_{0}$ is introduced from $i_{\rm CPL}$ to treat the nonlinear characteristic of CPL. 
To compute \eqref{eq:observables} at time $k$, we need to use the memory of past time series as in \eqref{eq:computeGSSA}, and hence we term \eqref{eq:observables} as the dynamic observable in the following. 
Note that the GSSA approach is used to derive a time-invariant model of the AC-DC converter in \cite{ijcta}: see Appendix~\ref{physical-model}. 

\subsection{Introduction to the Input Variable} \label{introduction-to-input}

Here, we introduce the input variable in discrete time $k$ that is determined in the K-MPC controller. 
As in Sec.~\ref{ac-dc-converter}, the duty ratio $\mu(t)\in[-1,1]$ is formulated as the control signal in continuous time $t$. 
The duty ratio can be represented as the sine wave called sinusoidal PWM (SPWM) \cite{spwm-introduction}, given by 
\begin{equation}
    \label{eq:mu}
    \mu(t)=u_1\sin\omega t+u_2\cos\omega t,
\end{equation}
where $u_1,u_2$ are constants at steady states. 
In this paper, we regard $u_1,u_2$ as the control variables that actuate the lifted dynamics $\boldsymbol{z}$ in discrete time. 
That is, their values at discrete time $k$ (namely, $\boldsymbol{u}[k]=[u_1[k],u_2[k]]^\top$) regulate the values of the lifted variable $\boldsymbol{z}$ at time $k+1$ ($\boldsymbol{z}[k+1]$) as in \eqref{eq:linear-representation}. 
This implies that $\mu(t)$ is updated at every period of the AC voltage. 
This requires synchronization in the experimental implementation, which is achieved using the phase-locked loop algorithm of \cite{yokoyama2009}. 

\subsection{Model Validation} \label{model-validation}

We validate the effectiveness of the dynamic observable in Sec.~\ref{synthesis-of-lifted-dynamics}. 
The estimated linear model with observables \eqref{eq:observables} is evaluated in terms of prediction performance. 
A collection of training and test datasets, which included $\mu(t),i(t),v(t)$ sampled under the period $\Delta t$, was obtained experimentally. 
The training dataset was generated using the input series as $u_{1}[k]=\bar{u}_{1}+\varDelta u_{1}$, $u_{2}[k]=\bar{u}_{2}+\varDelta u_{2}$, where $\bar{u}_{1}$, $\bar{u}_{2}$ are the nominal steady inputs given below, and $\Delta u_{1}$, $\Delta u_{2}$ are uniform random noise in $[-0.1,0.1]$. 
The nominal steady inputs, which are originally derived in \cite{ijcta} as $\mu(t)$ in continuous time $t$, are represented as $\bar{u}_{1} = -2(GV_{\rm d}+P/V_{\rm d})/I_{\rm d}$ and $\bar{u}_{2} = \omega LI_{\rm d}/V_{\rm d}$. 
The input series were experimentally realized with the PWM signal $s(t)$ determined by $\mu(t)$ in \eqref{eq:mu}. 
Then, the AC current $i(t)$ and the DC voltage $v(t)$ were measured with uniform sampling of $\Delta t=200\,\rm \mu s$ ($N=100$), yielding $40000$ samples of $(i(j\Delta t),v(j\Delta t))$. 
Thus, from (4), the $K=400$ samples of $(\langle i \rangle_h[k],\langle v \rangle_h[k])$ were derived and used for the KOM. 
In addition, the test dataset for the evaluation was obtained in the same way as the training data, under different input series due to the inclusion of random noise. 

\begin{figure}[t]
    \centering
    \includegraphics[width=1\linewidth]{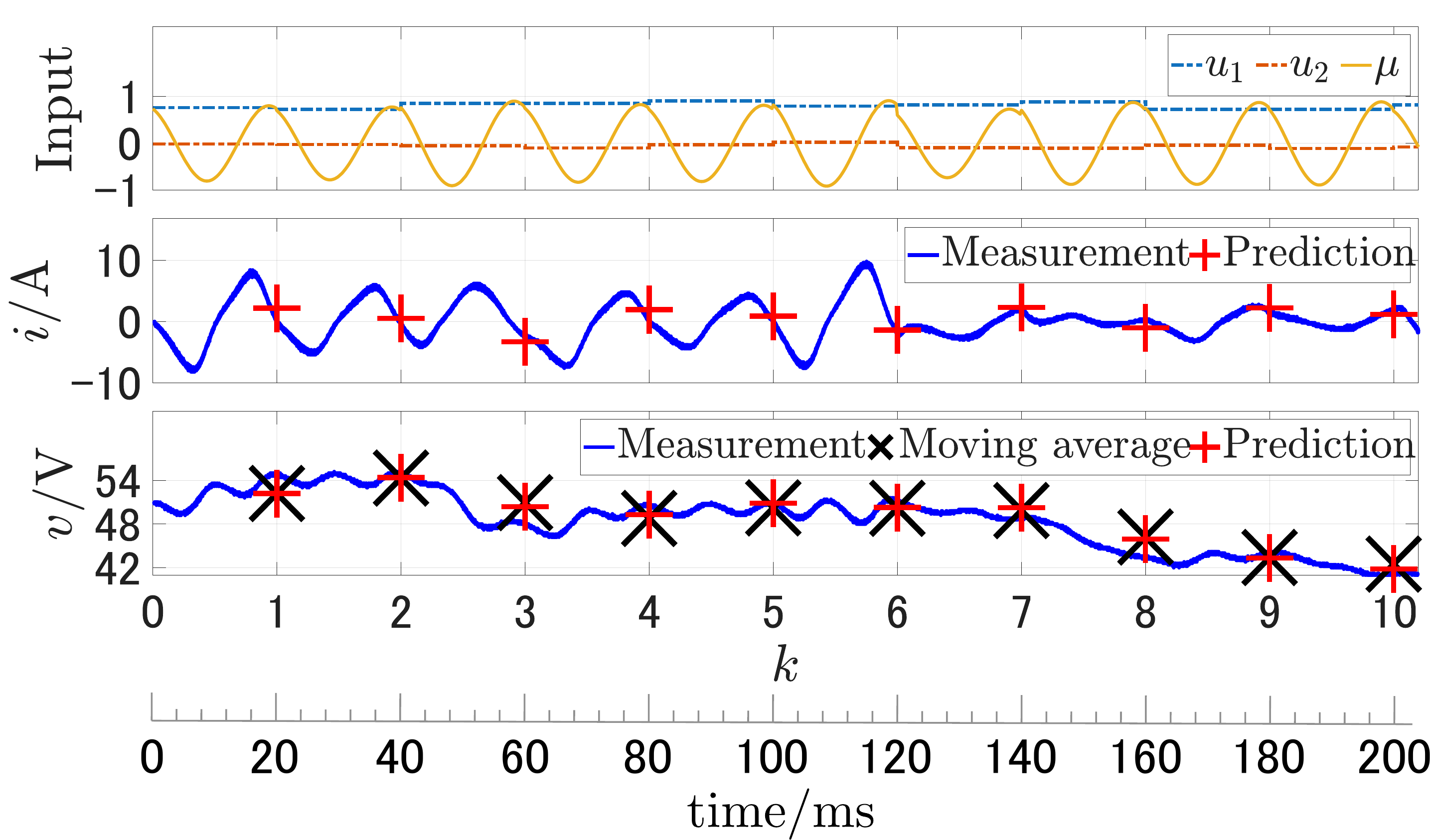}
    \caption{Comparison of measured data and predictions by the Koopman model.}
    \label{fig:modeling-verification}
\end{figure}

We compare the predictions of the GSSA $\langle i \rangle_{1}[k]$ and $\langle v \rangle_{0}[k]$ by the Koopman model with the measured data ($i(t),v(t)$). 
To do this, we computed the GSSA $\langle i \rangle_{1}[k],\langle v \rangle_{0}[k],\langle v^{-1} \rangle_{0}[k]$ for $k=1,\ldots,10$ from the $1000$ samples of test data $(i(j\varDelta t),v(j\varDelta t))$.
Then, deriving $\boldsymbol{z}[1]$ from $\langle i \rangle_{1}[1],\langle v \rangle_{0}[1],\langle v^{-1} \rangle_{0}[1]$, we predicted $\boldsymbol{z}[k]$ for $k=2,\ldots,10$ by the obtained Koopman model. 
Since $\langle v \rangle_{0}[k]$ is related to the mean value of $v(t)$, it is possible to compare $\langle v \rangle_{0}[k]$ with $v(t)$. 
Whereas, since $\langle i \rangle_{1}[k]$ is related to the phasor of $i(t)$, it is not possible to compare $\langle i \rangle_{1}[k]$ with the instantaneous value $i(t)$ directly. 
Thus, we calculated the prediction of the AC current $i(t)$ at time $t=kN\varDelta t$ by $\hat{i}(kN\varDelta t)\sim\langle i \rangle_{1}[k] \exp(\mathrm{j}\omega kN\varDelta t)+\langle i \rangle_{-1}[k] \exp(-\mathrm{j}\omega kN\varDelta t)$, where $\langle i \rangle_{1}[k]$ is the first element of $\boldsymbol{z}[k]$ and $\langle i \rangle_{-1}[k]$ is its conjugate. 
Here, it has been assumed that the AC current does not contain harmonics other than the first-order harmonic. 
Fig.~\ref{fig:modeling-verification} shows the comparison between the measured data and their predictions by the Koopman model. 
The figure in row 1 shows $\mu(t)$ (\emph{yellow} line), $u_{1}[k]$ (\emph{blue} line), and $u_{2}[k]$ (\emph{orange} line). 
The $u_1,u_2$ are plotted using the zeroth-order holding. 
The SPWM $\mu(t)$ changes at each of the discrete onsets $k=1,\ldots,10$. 
The figure in row 2 shows the measured $i(t)$ (\emph{blue} line) and $\hat{i}(kN\varDelta t)$ (\emph{red} $+$).
The prediction $\hat{i}(kN\varDelta t)$ is consistent with the measured current $i(t)$. 
The figure in row 3 shows the measured $v(t)$ (\emph{blue} line), $\langle v \rangle_{0}[k]$ from measured data (\emph{black} $\times$), and $\langle v \rangle_{0}[k]$ as the third element of $\boldsymbol{z}[k]$ (\emph{red} $+$). 
The prediction is consistent with $\langle v \rangle_{0}[k]$ computed from the measurement. 
It is worth noting that $\langle v \rangle_{0}[k]$ represents the mean value of $v(t)$ for $[(k-1)N\varDelta t,kN\varDelta t]$.
These results show that the estimated Koopman model captures the converter dynamics well. 
Therefore, we incorporate this Koopman model into the optimization problem of K-MPC. 

\subsection{MPC Formulation}

The K-MPC controller then solves at each time step $k$ of the closed-loop operation the following optimization problem: 
\begin{mini!}|s|
    {\substack{\boldsymbol{u}[k],\ldots,\\\boldsymbol{u}[k+N_{\rm P}-1]}}
    {\sum_{\ell=0}^{N_{\rm P}-1} \Big\{(\hat{\boldsymbol{z}}[\ell+1]-\boldsymbol{r})^{\top}\boldsymbol{Q}(\hat{\boldsymbol{z}}[\ell+1]-\boldsymbol{r})}{}{\notag}\breakObjective{+\varDelta\boldsymbol{u}[\ell]^{\top}\boldsymbol{R}\varDelta\boldsymbol{u}[\ell]\Big\}}
    {\label{eq:kmpc2}}
    {}
    \addConstraint{\hat{\boldsymbol{z}}[\ell+1]}{=\boldsymbol{A}\hat{\boldsymbol{z}}[\ell]+\boldsymbol{B}\boldsymbol{u}[k+\ell] \label{eq:kmpc2-state-transition}}
    \addConstraint{\hat{\boldsymbol{z}}[0]}{=\boldsymbol{z}[k] \label{eq:kmpc2-initial-condition}}
    \addConstraint{\varDelta\boldsymbol{u}[\ell]}{=\boldsymbol{u}[k+\ell] - \boldsymbol{u}[k+\ell-1] \label{eq:kmpc2-def-derivative-input}}
    \addConstraint{\bar{u}_{1}-0.1}{\leq u_{1}[k+\ell] \leq \bar{u}_{1}+0.1 \label{eq:kmpc2-input-constraint1}}
    \addConstraint{\bar{u}_{2}-0.1}{\leq u_{2}[k+\ell] \leq \bar{u}_{2}+0.1 \label{eq:kmpc2-input-constraint2}}
    \addConstraint{-1.8 }{\leq \hat{z}_{1}[\ell] \leq 0 \label{eq:kmpc2-const-z1}}
    \addConstraint{-0.87 }{\leq \hat{z}_{2}[\ell] \leq 0.87.\label{eq:kmpc2-const-z2}}
\end{mini!}
The integer $N_{\rm P}$ is the prediction horizon. 
We set $N_{\rm P}=3~(60\,\rm ms)$ based on the aircraft's electric standards \cite{MIL-STD-704F}, where the DC voltage is stipulated to restore in the order of several tens of milliseconds.
The cost function in \eqref{eq:kmpc2} is to penalize the deviation of the lifted variables $\boldsymbol{z}$ from the reference $\boldsymbol{r}$ and the increment of the input variables defined in \eqref{eq:kmpc2-def-derivative-input}, latter of which is motivated by minimizing the fluctuation of the input series in steady state. 
We set the reference $\boldsymbol{r}=[-I_{\rm d}/2,0,V_{\rm d},1/V_{\rm d}]^{\top}$ from Sec.~\ref{ac-dc-converter}. 
We also set the state weight matrix $\boldsymbol{Q}=\mathrm{diag}([0,1,1,0])$ because focusing only on $\Re\langle i \rangle_{1}$ and $\langle v \rangle_{0}$ \add{is} sufficient to achieving the control objectives in Sec.~\ref{ac-dc-converter}. 
The reference value of $\Im\langle i \rangle_{1}$, $-I_{\rm d}/2$, is not useful because, as seen in \eqref{eq:power-balance}, $I_{\rm d}$ is dependent on the parasitic resistance $r$, which is uncertain. 
The input weight matrix is set at $\boldsymbol{R}=0.1\boldsymbol{I}_{2}$ where $\boldsymbol{I}_{2}$ is the $2\times2$ identity matrix. 
The Koopman model is introduced as \eqref{eq:kmpc2-state-transition}, where $\hat{\boldsymbol{z}}[\ell]=[\hat{z}_1[\ell],\ldots,\hat{z}_4[\ell]]^\top$ denotes the prediction of $\boldsymbol{z}[k+\ell]$ ($\ell=1,\ldots,N_{\rm P}$). 
The initial value of $\hat{\boldsymbol{z}}$ is set as \eqref{eq:kmpc2-initial-condition}. 
The constraints \eqref{eq:kmpc2-input-constraint1} and \eqref{eq:kmpc2-input-constraint2} limit the input variables $u_1,u_2$ to the pre-defined range used to obtain the training dataset. 
The constraints \eqref{eq:kmpc2-const-z1} and \eqref{eq:kmpc2-const-z2} are related to the AC current limit in \eqref{eq:limit} and a minimum power factor to $0.9$ from the practical point of view \cite{blaabjerg2006}. 
When the AC current takes the maximum amplitude and the minimum power factor i.e., $i(t)=(4\,{\rm A})\sin(\omega t+\theta)$ with $ \cos\theta=0.9$, we have $\langle i \rangle_{1}=\pm0.87-\mathrm{j}1.8$, which is encoded in the constraints \eqref{eq:kmpc2-const-z1} and \eqref{eq:kmpc2-const-z2}. 
\add{In the experiment, we use MATLAB function `quadprog' to solve the optimization problem (7). }

\begin{figure}[t]
    \centering
    \includegraphics[width=0.77\linewidth]{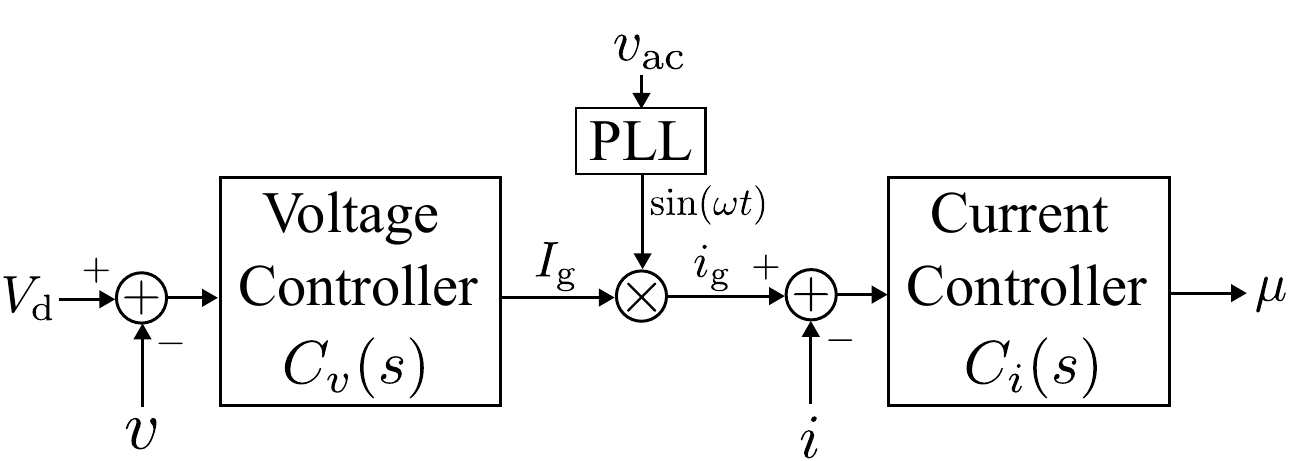}
    \caption{Block diagram of the cascade PI controller.}
    \label{fig:PIcontroller}
\end{figure}

\begin{figure*}[t]
    \begin{center}
    \begin{tabular}{ccc}
        \begin{minipage}{0.31\linewidth}
            \begin{center}
                \includegraphics[scale=0.15]{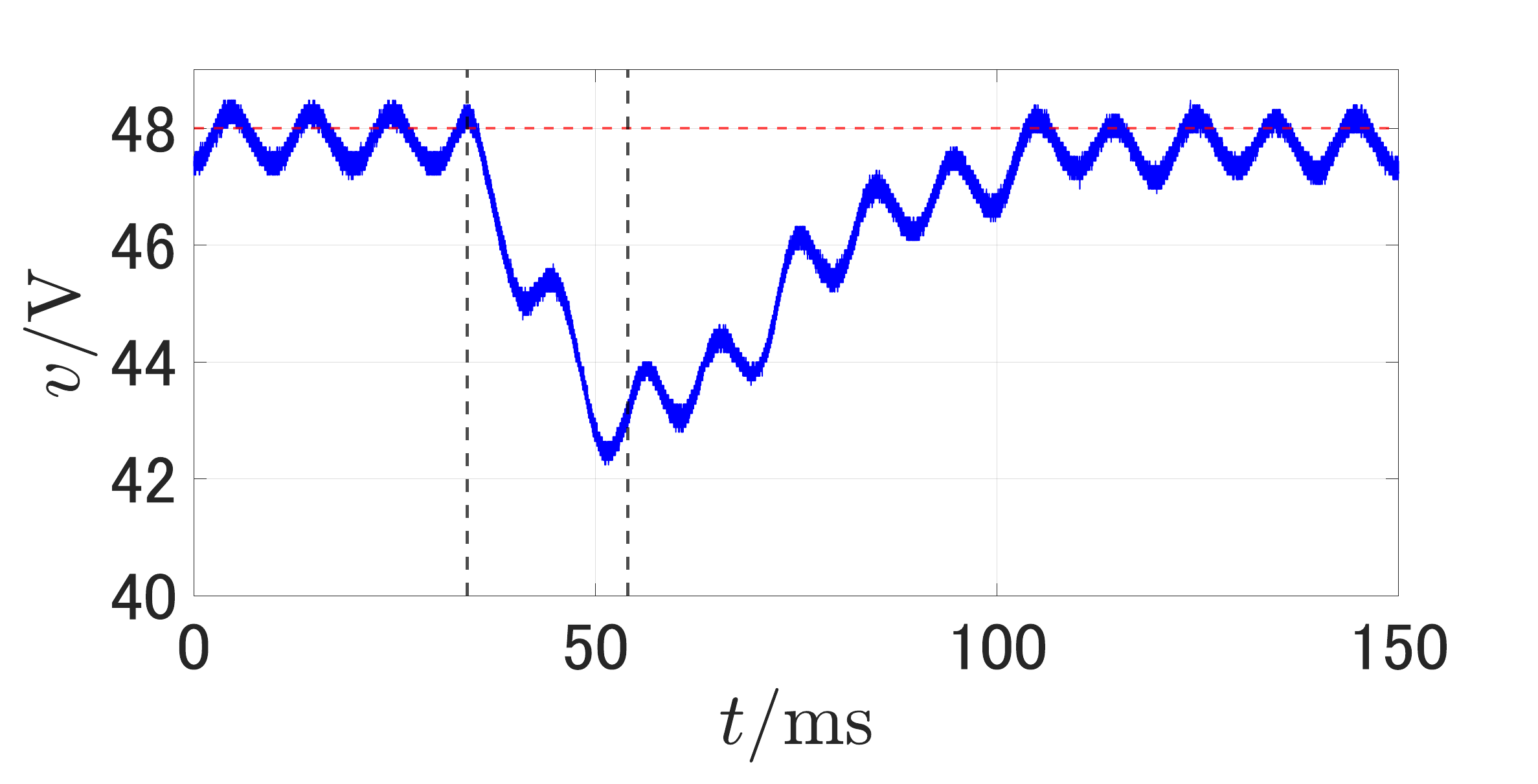}
            \end{center}
        \end{minipage} &
        \begin{minipage}{0.31\linewidth}
            \begin{center}
                \includegraphics[scale=0.15]{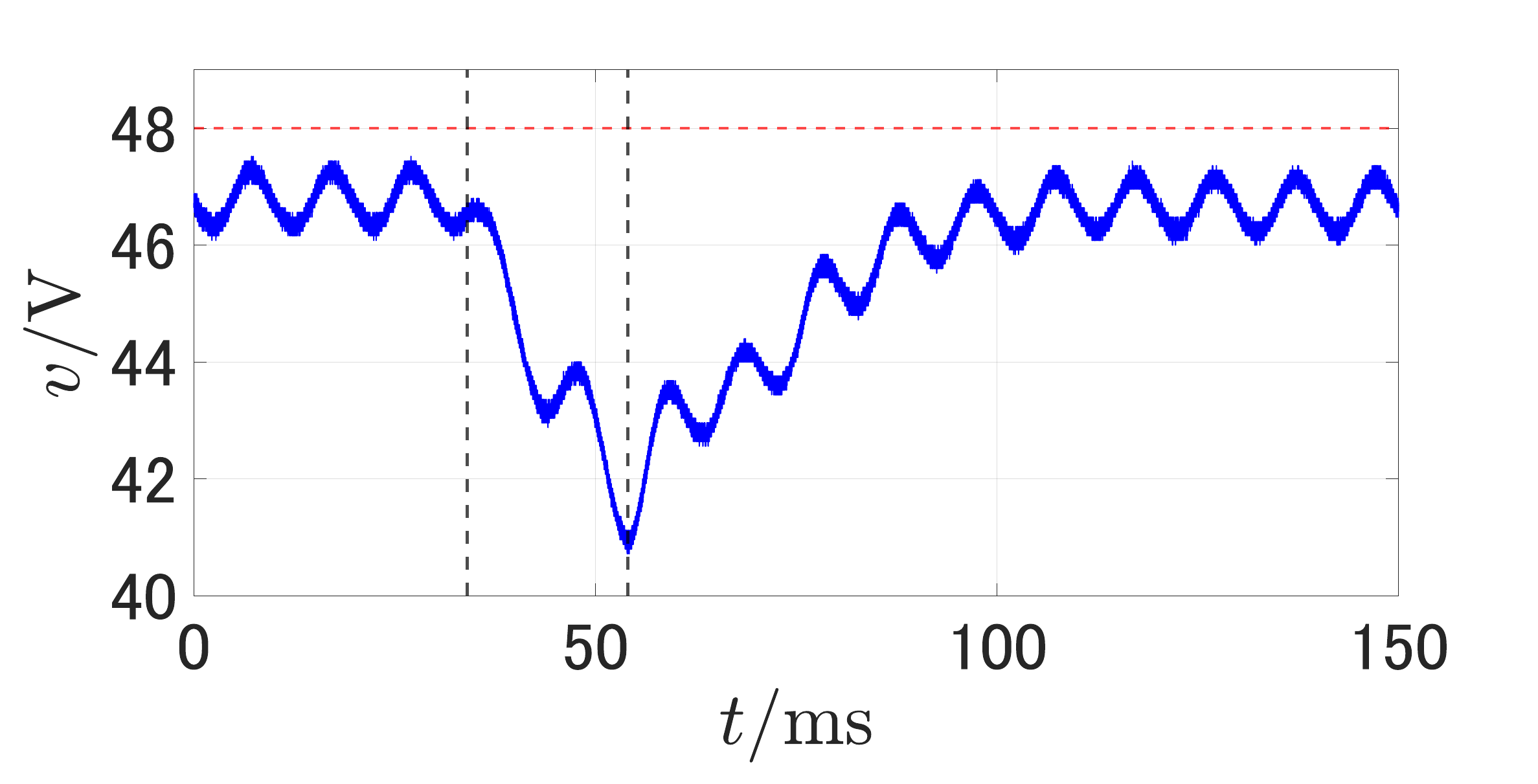}
            \end{center}
        \end{minipage} &
        \begin{minipage}{0.31\linewidth}
            \begin{center}
                \includegraphics[scale=0.15]{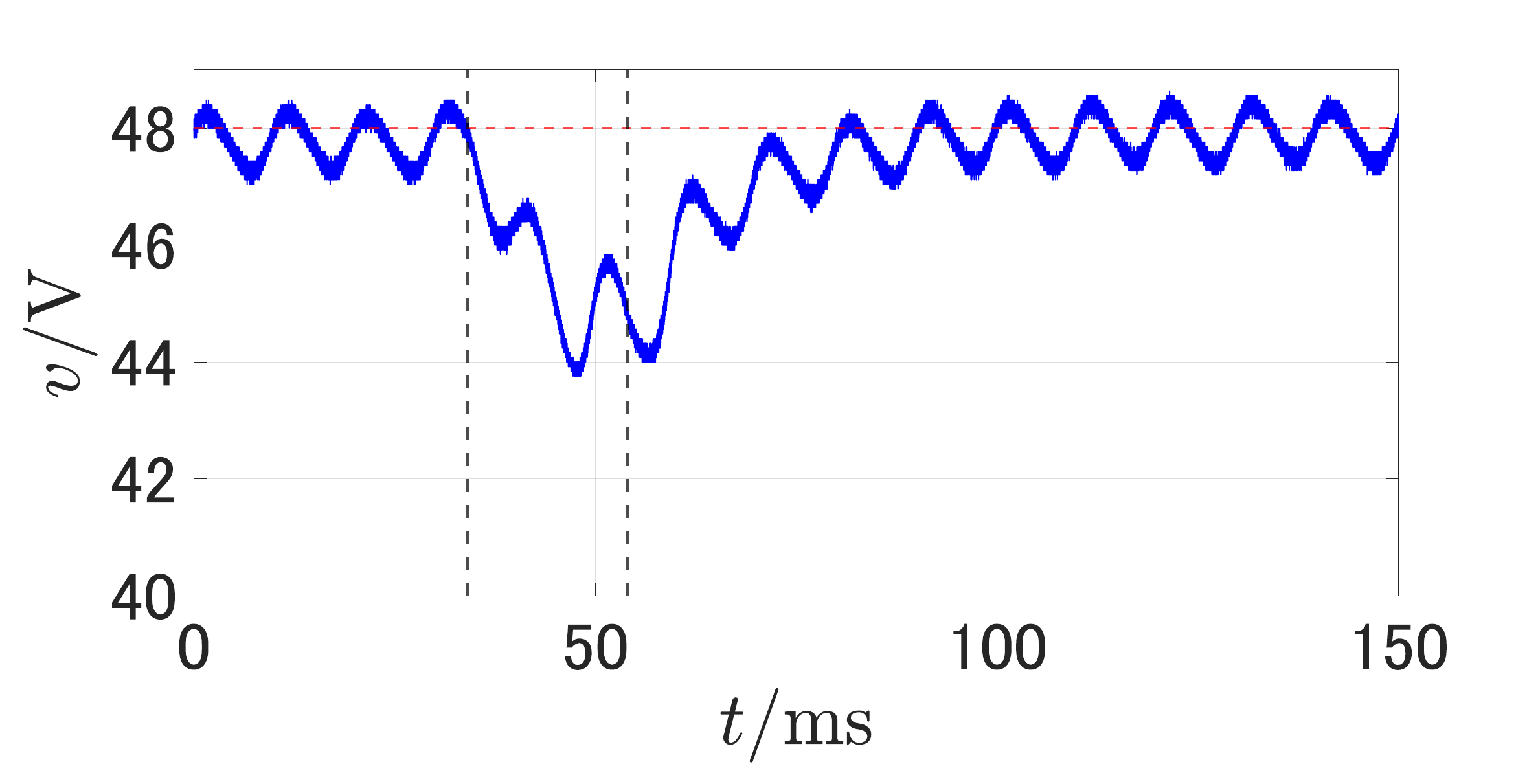}
            \end{center}
        \end{minipage} 
        \\
        \begin{minipage}{0.31\linewidth}
            \begin{center}
                \includegraphics[scale=0.15]{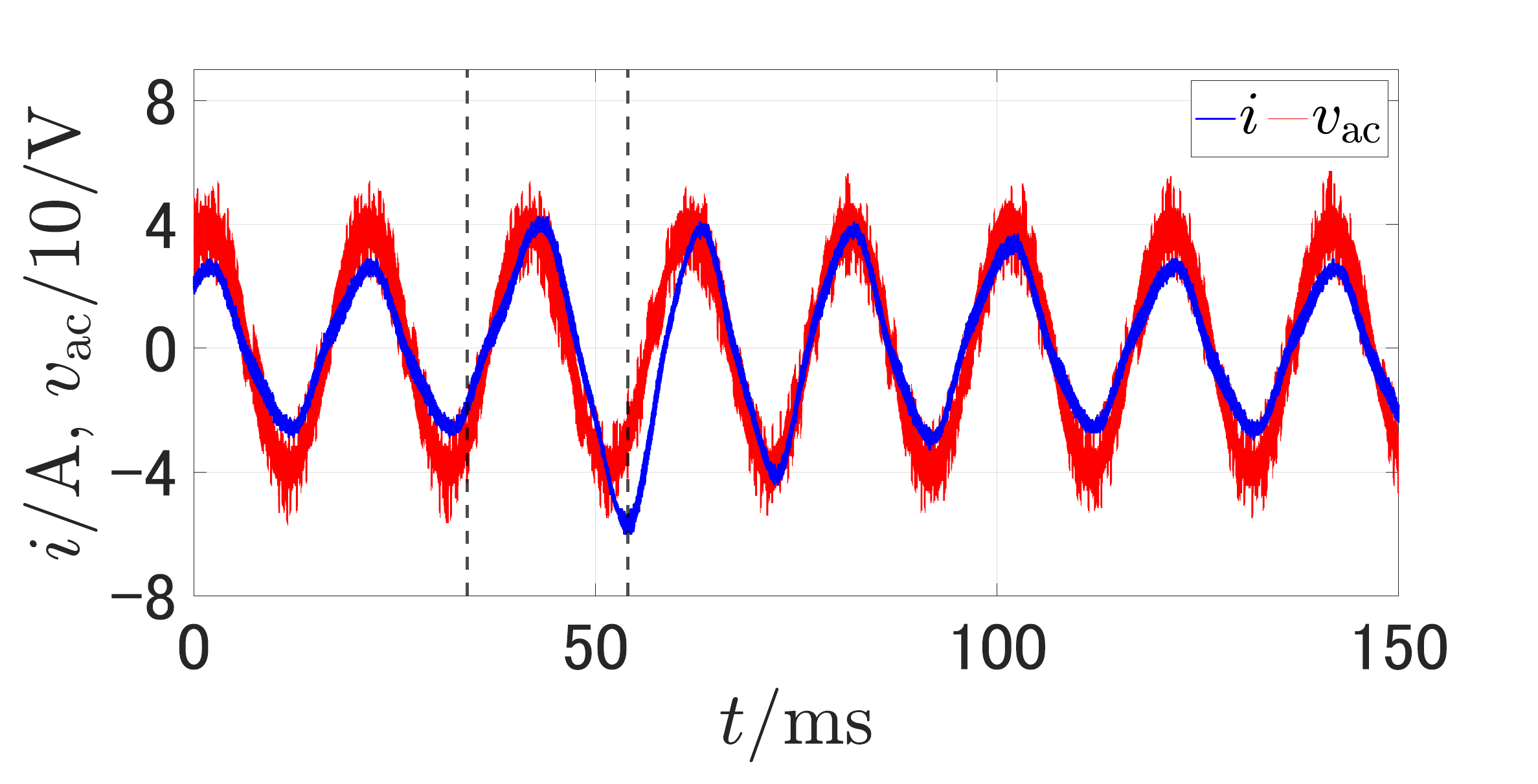}
            \end{center}
        \end{minipage} &
        \begin{minipage}{0.31\linewidth}
            \begin{center}
                \includegraphics[scale=0.15]{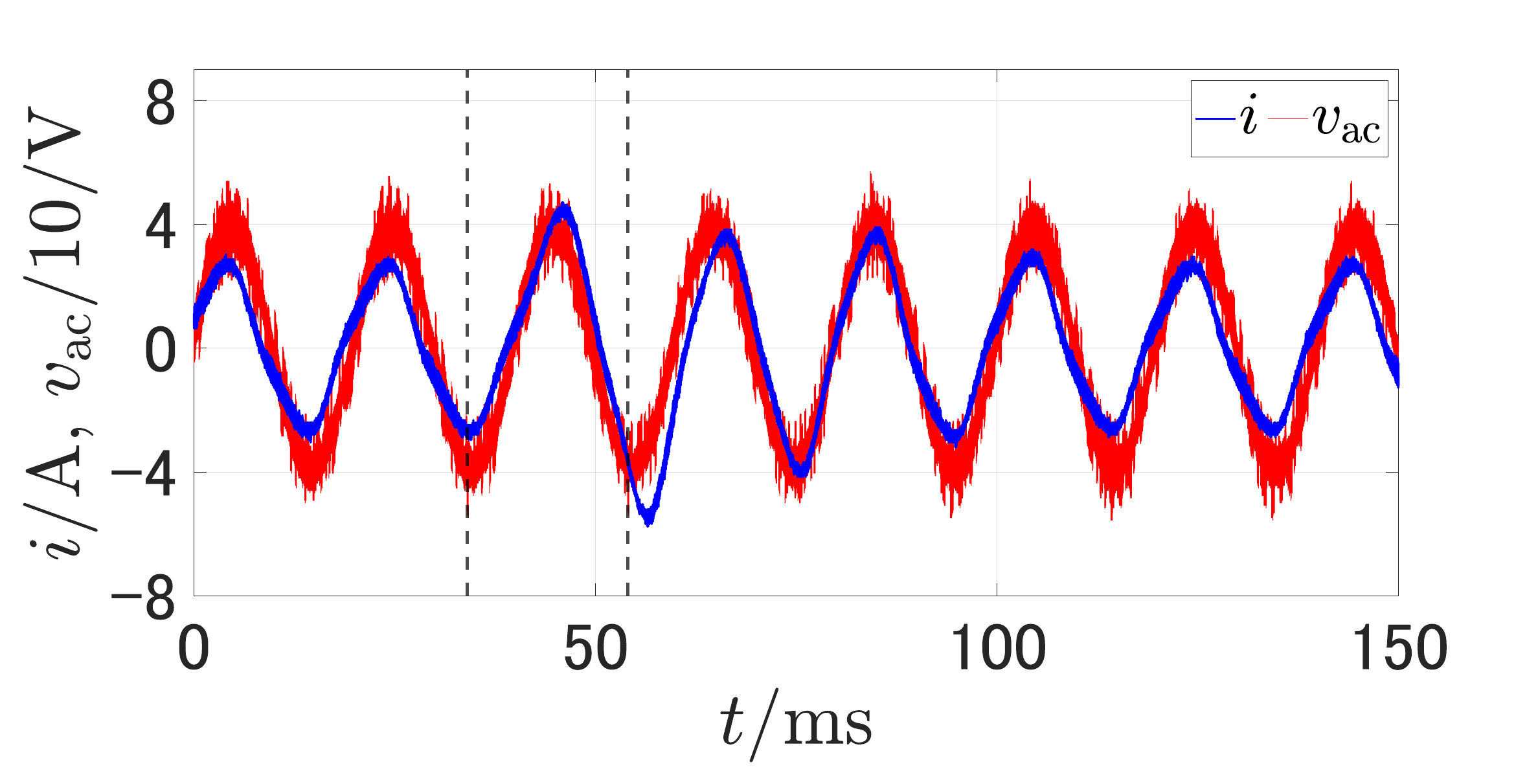}
            \end{center}
        \end{minipage} &
        \begin{minipage}{0.31\linewidth}
            \begin{center}
                \includegraphics[scale=0.15]{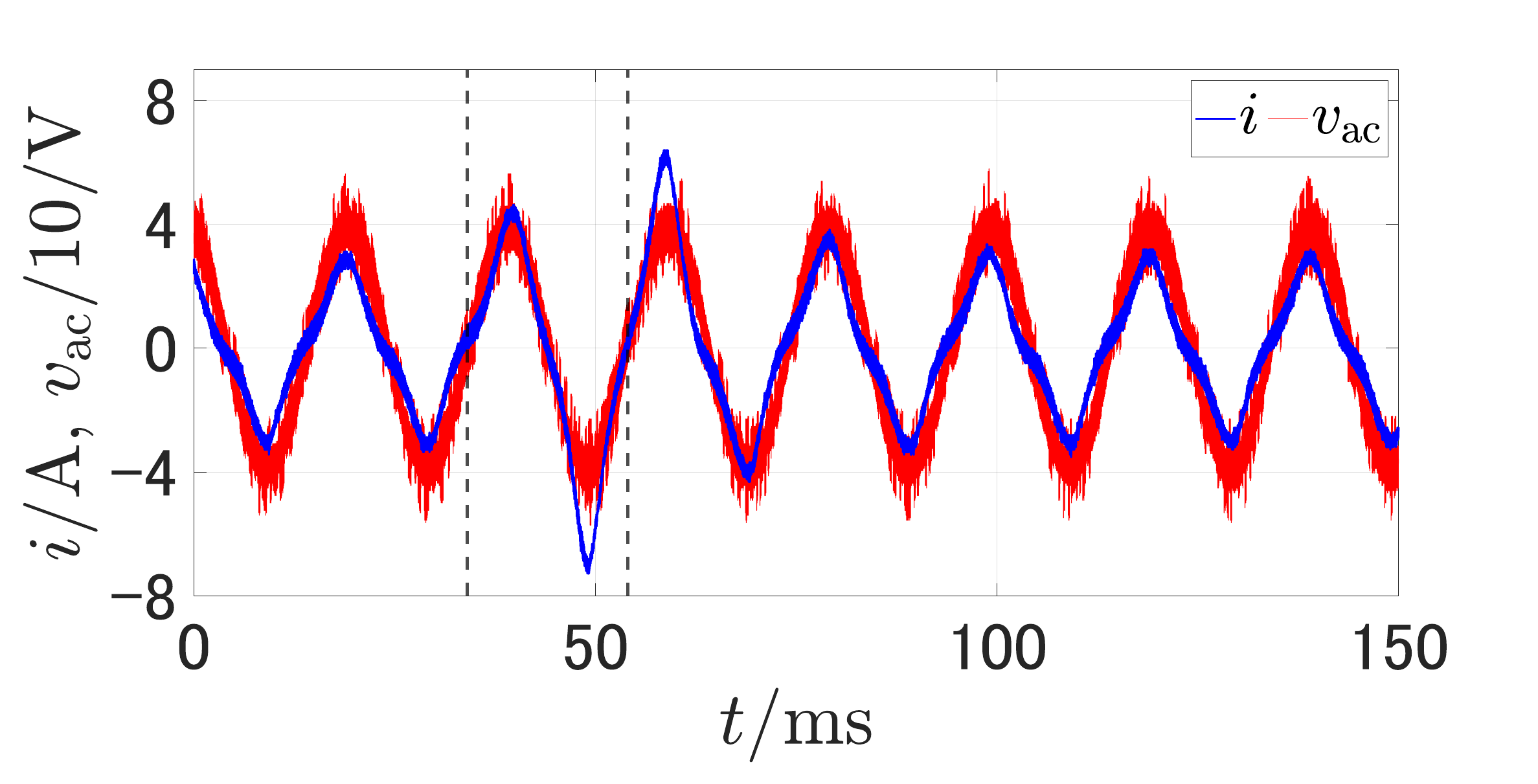}
            \end{center}
        \end{minipage}
        \\
        \begin{minipage}{0.31\linewidth}
            \begin{center}
                \includegraphics[scale=0.15]{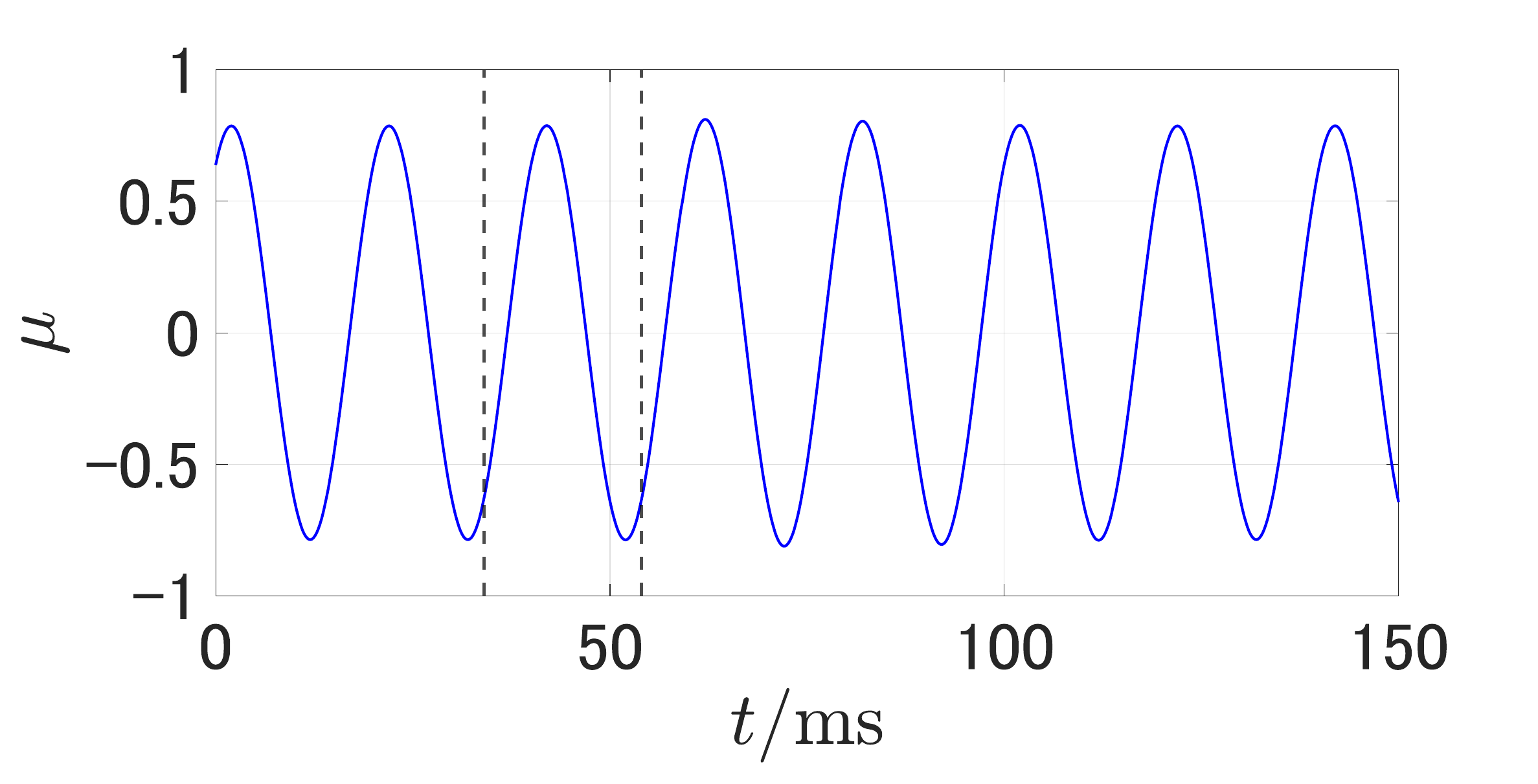}
            \end{center}
        \end{minipage} &
        \begin{minipage}{0.31\linewidth}
            \begin{center}
                \includegraphics[scale=0.15]{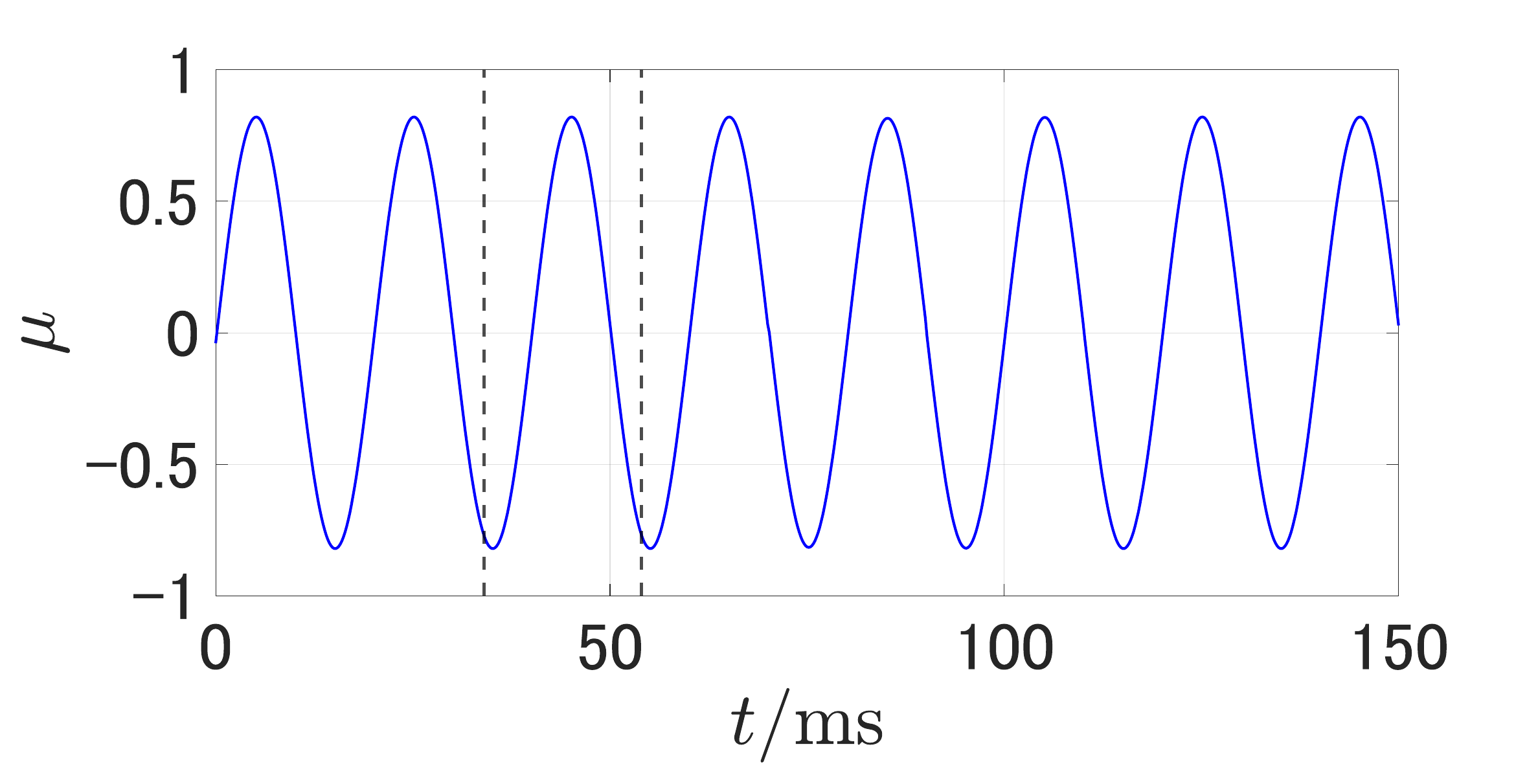}
            \end{center}
        \end{minipage} &
        \begin{minipage}{0.31\linewidth}
            \begin{center}
                \includegraphics[scale=0.15]{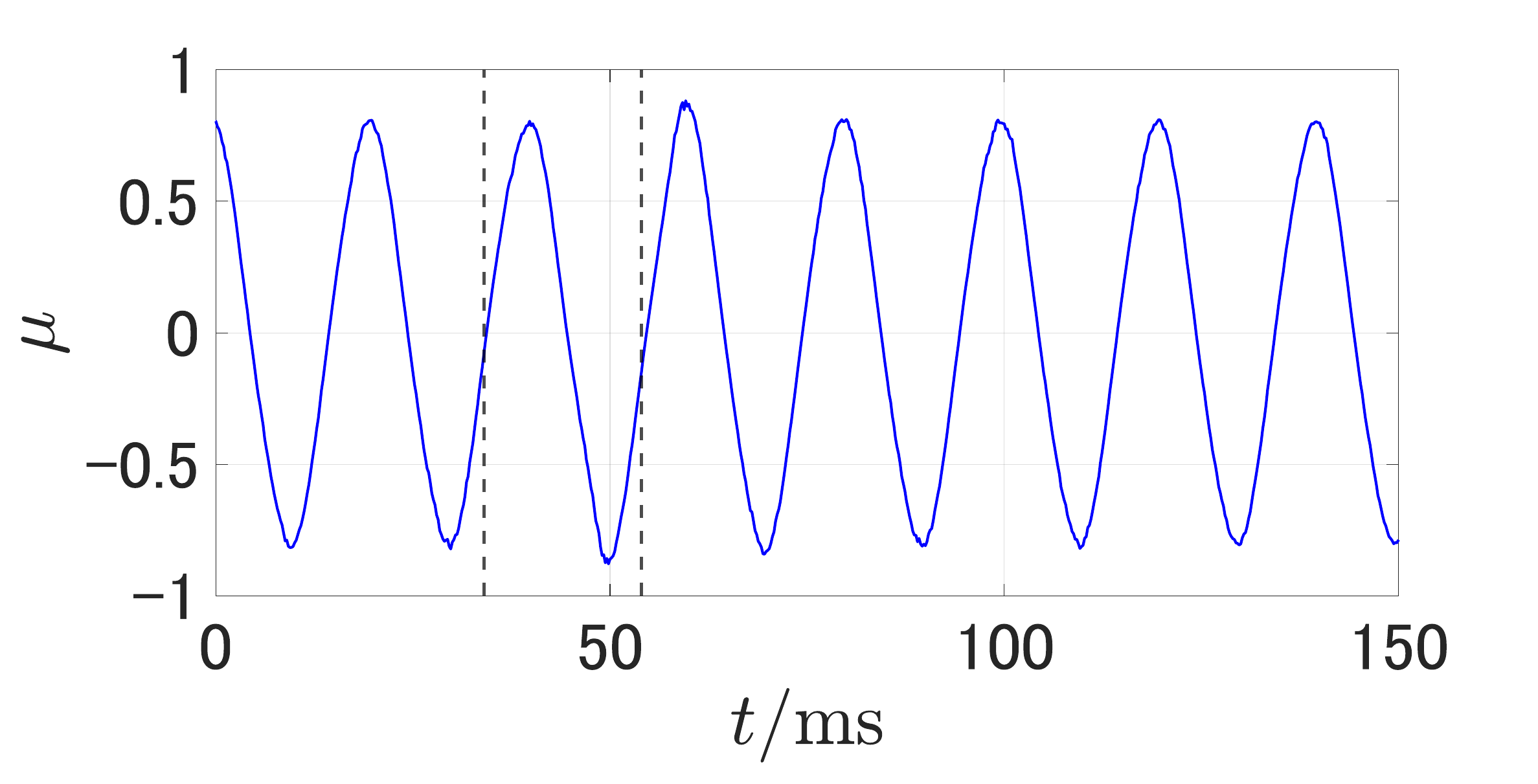}
            \end{center}
        \end{minipage}
    \end{tabular}
    \caption{Experimental results of K-MPC (left), IDA-PBC (center), and PI (right). 
    Graphs in row 1 show the DC voltage (\emph{blue} line) and its reference (\emph{red} line), graphs in row 2 the AC current (\emph{blue} line) and the AC supply voltage (\emph{red} line), and graphs in row 3 the duty ratio. 
    The vertical \emph{dotted} lines represent $34\,\rm ms$ and $54\,\rm ms$, respectively.}
    \label{fig:transient}
    \end{center}
\end{figure*}

\section{Experimental Results}\label{experimental-result}


\subsection{Conventional Strategies}

We use two conventional control strategies for the AC-DC converter. 
The first is based on \cite{ijcta}, where the authors applied the Interconnection and Damping Assignment-Passivity-Based Control (IDA-PBC) \cite{ORTEGA2002585}. 
The control input is given in the form of \eqref{eq:mu}. 
The authors of \cite{ijcta} use the DC current flowing into DC loads, denoted by $i_{\ell}$, for their feedback control and give the formula $u_{1}=-2i_{\ell}/I_{\rm d}$ and $u_{2}=\omega LI_{\rm d}/V_{\rm d}$. 
Therefore, we measure the DC current flowing into the resistive load and the CPL, and use its moving average over AC period as $i_{\ell}$
, thereby implementing their feedback control. 
\add{We use the values described in Sec.~\ref{ac-dc-converter} for $L$ and $I_{\rm d}$ to compute $u_1,u_2$.}

The second is the PI control, widely used in practical power converters to regulate both $i$ and $v$ \cite{Euzeli}. 
Fig.~\ref{fig:PIcontroller} shows the brief structure of the cascade PI controller used in our experiment, where the PI controller $C_{v}(s) = K_{v,\rm p} + K_{v,\rm i}/s$ and the Proportional Resonant (PR) controller $C_{i}(s) = K_{i,\rm p} + 2K_{i,\rm r}\omega_{c}s/(s^{2}+2\omega_{c}s+\omega^{2}_{0})$ in \cite{PRcontroller} are introduced with constant parameters $K_{v,\rm p},K_{v,\rm i},K_{i,\rm P}, K_{i,{\rm R}},\omega_c,\omega_0$. 
The PI controller outputs the reference amplitude $I_{\rm g}$ of the AC current $i$. 
Then, the instantaneous current reference $i_{\rm g}$ is generated by multiplying $I_{\rm g}$ with the sinusoidal wave of the estimated phase via the PLL algorithm in Sec.~\ref{introduction-to-input}. 
The parameter $\omega_0, \omega_c$ was set to $50\,\rm Hz,0.01\omega_0$ in this paper. 
The gains were set to $K_{\rm v,p}=1$, $K_{\rm v,i}=2$ for the PI controller and $K_{\rm i,p}=0.04$, $K_{\rm i,r}=4$ for the PR controller. 
We obtained rough values of 
the gains of the PI and PR controllers based on the symmetrical optimum method \cite{outerPItuning} and  \cite{PRtuning}, respectively. 
Then, we manually tuned the rough values to achieve better performance in mitigating overshoot. 

\subsection{Results and Discussion}

Figure~\ref{fig:transient} shows the results of the control experiments of the AC-DC converter using the three different strategies: K-MPC (left), IDA-PBC (center), and PI control (right). 
The set-point of CPL was initially set at $25\,\rm W$, while it was temporally increased to $100\,\rm W$ during $[34,54]\,\rm ms$, denoted by the vertical \emph{dotted} lines. 
Regarding the control objective 1), K-MPC and PI control achieve a smaller steady-state error 
for the mean value of the DC voltage $v$, while IDA-PBC has a non-zero error of about $1.5\,\rm V$. 
Regarding the control objective 2), the power factors measured with the power analyzer (PPA5530 from IWATSU ELECTRIC CO., LTD.) were $0.99$ for K-MPC, $0.93$ for IDA-PBC, and $0.97$ for PI control. 
The performance of IDA-PBC might be degraded due to parameter uncertainty, while K-MPC and PI control address it by the data-driven modeling and model free nature, respectively. 
These indicate that K-MPC and PI achieve better steady-state error performance. 

\begin{figure}
    \begin{center}
    \begin{tabular}{cc}
        \begin{minipage}{0.435\linewidth}
            \begin{center}
                \includegraphics[scale=0.26]{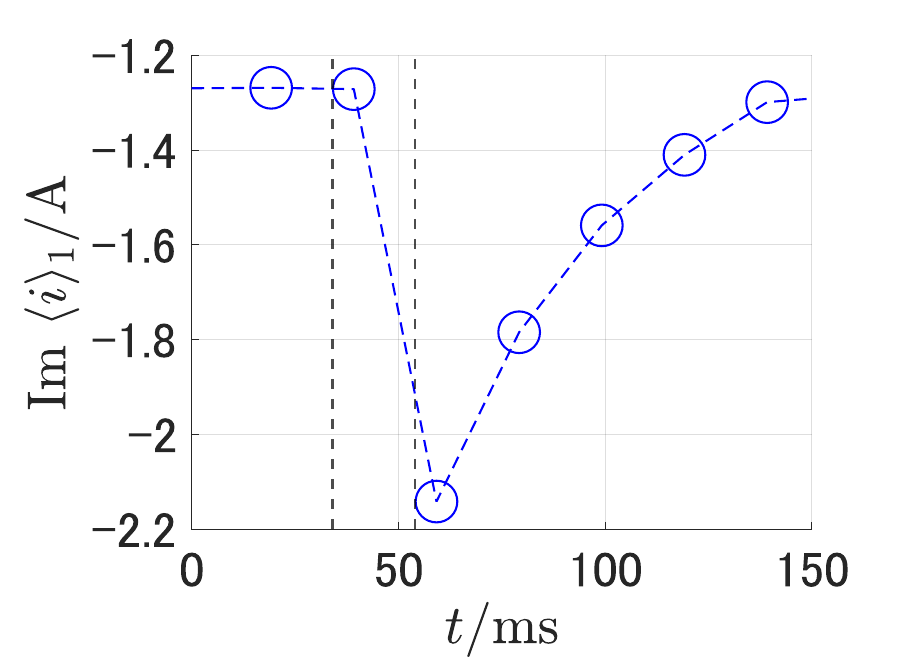}
                \subcaption{$\Im\langle i \rangle_{1}$}
            \end{center}
        \end{minipage} &
        \begin{minipage}{0.435\linewidth}
            \begin{center}
                \includegraphics[scale=0.26]{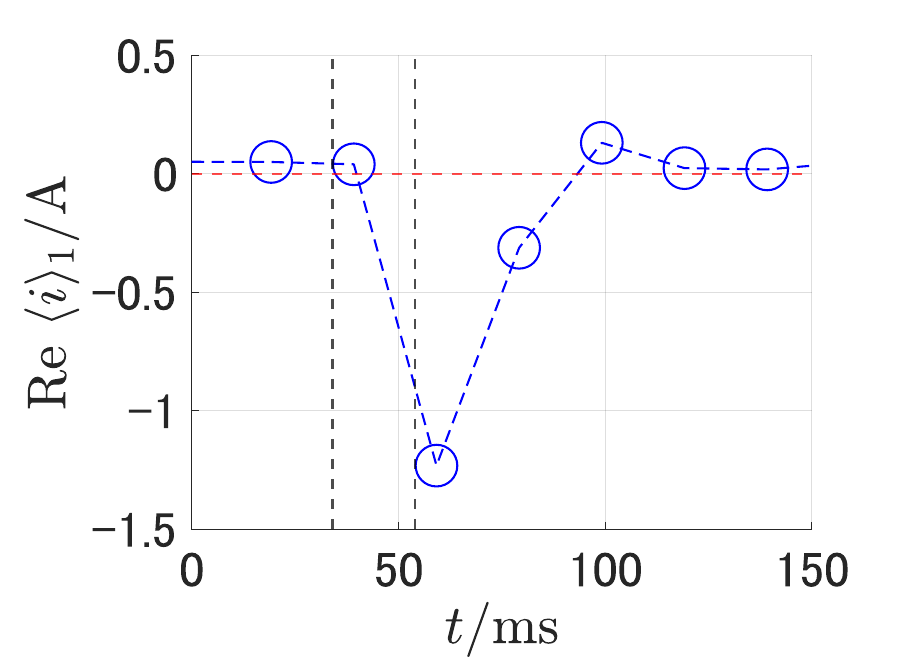}
                \subcaption{$\Re\langle i \rangle_{1}$}
            \end{center}
        \end{minipage}
        \\
        \begin{minipage}{0.435\linewidth}
            \begin{center}
                \includegraphics[scale=0.26]{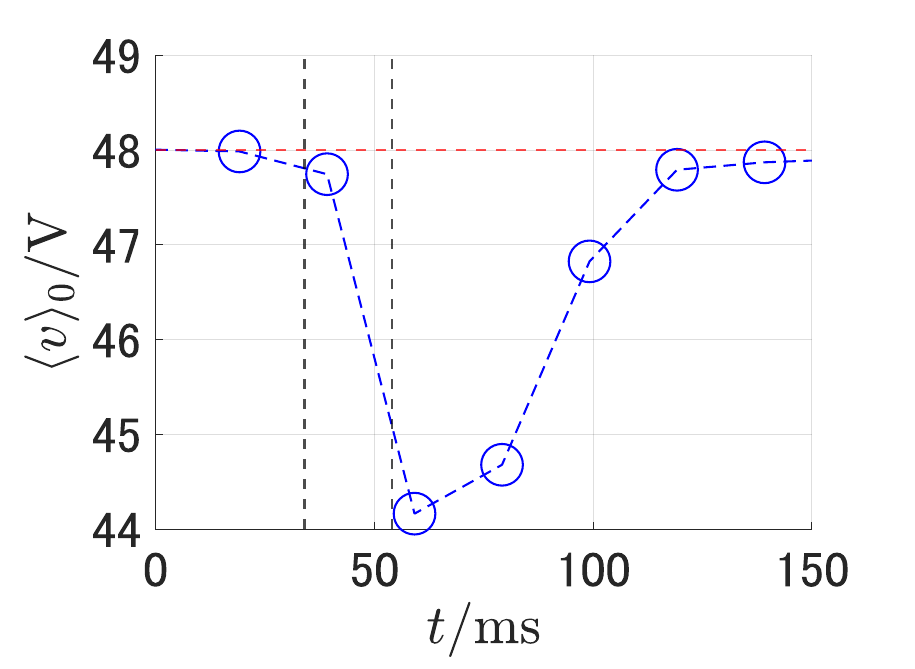}
                \subcaption{$\langle v \rangle_{0}$}
            \end{center}
        \end{minipage} &
        \begin{minipage}{0.435\linewidth}
            \begin{center}
                \includegraphics[scale=0.281]{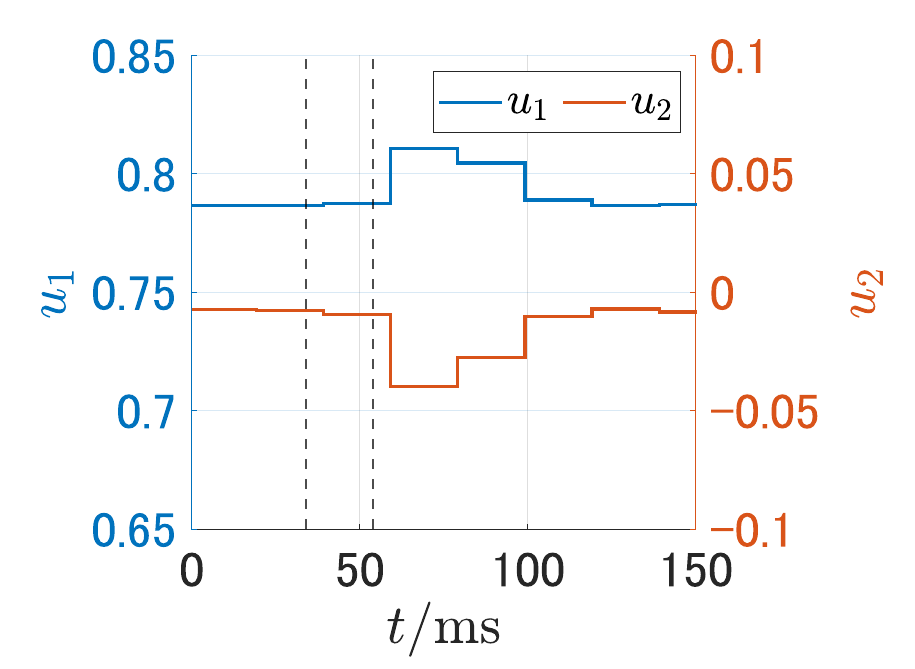}
                \subcaption{$\boldsymbol{u}$}
            \end{center}
        \end{minipage}
    \end{tabular}
    \caption{Transient responses of averaged variables and inputs in the K-MPC controller.}
    \label{fig:transient-gssa}
    \end{center}
\end{figure}

We next address the transient performance following the change of set point. 
As the first row, the DC voltage starts to drop at time $34\,\rm ms$ and to rise at $54\,\rm ms$.  
K-MPC and IDA-PBC require approximately $50\,\rm ms$ to restore the DC voltage, whereas PI control requires approximately $30\,\rm ms$. 
Here, K-MPC successfully limits the AC current as specified by the constraints after $54\,\rm ms$, while PI control allows for a large amplitude of the AC current. 
To see this, we plot the transient responses of the lifted variables $\boldsymbol{z}$ and the input ones $\boldsymbol{u}$ in Fig.~\ref{fig:transient-gssa}. 
The variables $\Im\langle i \rangle_{1}$, $\Re\langle i \rangle_{1}$, and $\langle v \rangle_{0}$ are plotted as the \emph{blue} circles in (a), (b), and (c), respectively. 
The \emph{red} lines in (b) and (c) represent the references. 
Since we have not imposed the weight on $\Im\langle i \rangle_{1}$, its reference is not described in (a). 
The input variables $\boldsymbol{u}$ are shown in (d), where  the \emph{blue} (or \emph{orange}) line represents $u_1$ (or $u_2$). 
The vertical \emph{dotted} lines represent $34\,\rm ms$ and $54\,\rm ms$. 
The variables $\Im\langle i \rangle_{1}$ and $\Re\langle i \rangle_{1}$ are within the constraints \eqref{eq:kmpc2-const-z1} and \eqref{eq:kmpc2-const-z2} except for one step right after the load change. 
This is because the CPL's change does not affect the averaged states at the beginning, and the K-MPC controller does not generate an input that satisfies the constraints \emph{during the change}. 
However, the K-MPC controller suppresses the AC current as soon as it detects the CPL's change from the variation in $\Im\langle i \rangle_{1}$ and $\Re\langle i \rangle_{1}$. 
This clearly shows that K-MPC can handle the hard constraints of the AC-DC converter, which is never achieved with PI control.

\section{Concluding Remarks}\label{conclusion}

This paper presents the first experimental demonstration of K-MPC for a real AC-DC converter. 
We estimated a linear time-invariant model of a nonlinear periodic-time-variant conversion dynamics from measured data and designed the K-MPC controller by formulating the optimization problem. 
We showed that the K-MPC controller achieved better steady-state and transient performance \add{with respect to the safety constraint} than IDA-PBC and PI control. 
Future work involves implementing online modeling and control to handle dynamic load changes more precisely. 

\appendix

\subsection{Physics-Based Model} \label{physical-model}
The dynamical model of the AC-DC converter in Fig.~\ref{cir:1} is represented by
\begin{equation}
    \label{eq:AC-DC}
    \left.
    \begin{alignedat}{2}
        -L\frac{\mathrm{d}i(t)}{\mathrm{d}t} &= s(t)v(t) + ri(t) - E\sin{\omega t} \\
        C \frac{\mathrm{d}v(t)}{\mathrm{d}t} &= s(t)i(t) - Gv(t) - \frac{P}{v(t)} \\
    \end{alignedat}
    \right\}.
\end{equation}
This model is clearly nonlinear and time-variant. 
Since the switching frequency $20\,\rm kHz$ is much higher than the AC frequency $50\,\rm Hz$, $s(t)\in\{-1,1\}$ can be approximately replaced with $\mu(t)\in[-1,1]$. 
Here, we introduce the continuous-time formulation of \eqref{eq:computeGSSA} as
\begin{equation}
    \label{eq:computeGSSA_c}
    \langle y \rangle_{h}(t) := \frac{1}{T} \int_{t-T}^{t} y(\tau)\exp({-\mathrm{j}\omega h\tau}){\rm d}\tau.
\end{equation}
From the properties $\mathrm{d}\langle y \rangle_{h}/\mathrm{d}t = \left\langle\mathrm{d}y/\mathrm{d}t \right\rangle_{h} - \mathrm{j}h\omega\langle y \rangle_{h}$ and $\langle xy \rangle_{h} = \sum^{\infty}_{g=-\infty} \langle x \rangle_{h-g} \langle y \rangle_{g}$ in \cite{gssa-introduction} and the assumption that $i(t)$ does not contain harmonics other than the first-order harmonic and $v(t)$ does not contain \add{harmonics whose amplitude is larger than the DC bias}, it is possible to rewrite \eqref{eq:AC-DC} to the nonlinear yet time-invariant model as
\begin{equation}
    \label{eq:AC-DC-GSSA}
    \left.
    \begin{alignedat}{2}
        -L\frac{\mathrm{d}\langle i \rangle_{1}}{\mathrm{d}t} &= w \langle v \rangle_{0} + r\langle i \rangle_{1} + \mathrm{j}\omega L\langle i \rangle_{1} + \mathrm{j}\frac{E}{2} \\
        C \frac{\mathrm{d}\langle v \rangle_{0}}{\mathrm{d}t} &= w \langle i \rangle_{-1} + w^{*} \langle i \rangle_{1} - G\langle v \rangle_{0} - \left\langle \frac{P}{v} \right\rangle_{0} \\
        w :=& u_{2}/2-\mathrm{j}u_{1}/2,~w^{*}:= u_{2}/2+\mathrm{j}u_{1}/2
    \end{alignedat}
    \right\}.
\end{equation}
\add{Note that \eqref{eq:AC-DC-GSSA} cannot be used as a prediction model for the MPC controller due to uncertainty of $P$ as desribed in Sec.~\ref{ac-dc-converter}.}

\section*{Acknowledgment}
This work was supported in part by JSPS KAKENHI (Grant No. 23H01434) and JSPS Bilateral Collaborations (Grant No. JPJSBP120242202).

\raggedbottom               
\IEEEtriggeratref{2}


\begin{thebibliography}{99}

\bibitem{IEEE-Electrific-Mag}
 ``Evolving electrification technology," {\em IEEE Electrific. Mag.}, vol.~11, no.~2, pp.~1--100, 2023.
 
\bibitem{IEEE-Power-Electron-Mag}
 ``Real-time simulator," {\em IEEE Power Electron. Mag.}, vol.~11, no.~3, pp.~1--132, 2024.

\bibitem{Euzeli}
E.~C. dos.~Santos. Jr and E.~R.~C. da~Silva, ``Control strategies for power converters", in {\em Advanced Power Electronics Converters}. Hoboken, NJ, USA: Wiley, 2014, ch.~9, pp.~264--294.

\bibitem{ijcta}
C.~Batlle, A.~Dòria-Cerezo, and E.~Fossas, ``Bidirectional power flow control of a power converter using passive Hamiltonian techniques,'' {\em Int. J. Circuit Theory Appl.}, vol.~36, no.~7, pp.~769--788, 2008.

\bibitem{mpc-for-converter}
S.~Vazquez, J.~Rodriguez, M.~Rivera, L.~G. Franquelo, and M.~Norambuena, ``Model predictive control for power converters and drives: Advances and trends,'' {\em IEEE Trans. Ind. Electron.}, vol.~64, no.~2, pp.~935--947, 2017.

\bibitem{onboard}
G.~Buticchi, S.~Bozhko, M.~Liserre, P.~Wheeler, and K.~Al-Haddad, ``On-board microgrids for the more electric aircraft—technology review,'' {\em IEEE Trans. Ind. Electron.}, vol.~66, no.~7, pp.~5588--5599, 2019.

\bibitem{barzkar}
A.~Barzkar and M.~Ghassemi, ``Electric power systems in more and all electric aircraft: A review,'' {\em IEEE Access}, vol.~8, pp.~169314--169332, 2020.


\bibitem{KORDA2018149}
M.~Korda and I.~Mezić, ``Linear predictors for nonlinear dynamical systems: Koopman operator meets model predictive control,'' {\em Automatica}, vol.~93, pp.~149--160, 2018.

\bibitem{koopman-introduction-springer}
A.~Mauroy, I.~Mezi\'{c}, and Y.~Susuki, {\em The Koopman Operator in Systems and Control: Concepts, Methodologies, and Applications}. Cham, Switzerland: Springer, 2020.

\bibitem{hanke2019}
S.~Hanke, S.~Peitz, O.~Wallscheid, S.~Klus, J.~Böcker, and M.~Dellnitz, ``Koopman operator-based finite-control-set model predictive control for electrical drives,'' 2019,
\textit{arXiv:1804.00854}.

\bibitem{koopman-dcdcconverter}
A.~Maksakov and S.~Palis, ``Koopman-based optimal control of boost DC-DC converter,'' in {\em Proc 2020 IEEE Problems of Automated Electrodrive. Theory and Pract. (PAEP)}, pp.~1--4.

\bibitem{Debnath2024}
R.~Debnath, D.~Kumar, G.~S. Gupta, S.~R. Samantaray, and E.~F. El-Saadany, ``A unified novel Koopman-based model predictive control scheme to achieve seamless stabilization of nonlinear dynamic transitions in inverter-based stochastic microgrid clusters,'' {\em IEEE Trans. Smart Grid}, vol.~15, no.~6, pp.~5441--5458, 2024.

\bibitem{HUO2025106225}
D.~Huo, A.~Adunyah, and C.~M. Hall, ``Real-time application of Koopman-based optimal control strategies for fuel cell stack thermal management,'' {\em Control Eng. Pract.}, vol.~156, Art.~no.~106225, 2025.

\bibitem{gssa-introduction}
S.~Sanders, J.~Noworolski, X.~Liu, and G.~Verghese, ``Generalized averaging method for power conversion circuits,'' {\em IEEE Trans. Power Electron.}, vol.~6, no.~2, pp.~251--259, 1991.

\bibitem{edmd-introduction}
M.~O.~Williams, I.~G.~Kevrekidis, and C.~W.~Rowley, ``Data-Driven approximation of the Koopman operator: Extending dynamic mode decomposition,'' {\em J. Nonlinear Sci}, pp.~1307--1346, 2015.

\bibitem{Brunton_Kutz_2022}
S.~L. Brunton and J.~N. Kutz, ``Data-driven dynamical systems'', in {\em Data-driven Science and Engineering: Machine Learning, Dynamical Systems, and Control}, 2nd ed. Cambridge, U.K., Cambridge Univ. Press, 2022, ch.~7, pp.~253--310.

\bibitem{Peyghami2020}
S.~Peyghami, P.~Palensky, and F.~Blaabjerg, ``An overview on the reliability of modern power electronic based power systems,'' {\em IEEE Open J. Power Electron.}, vol.~1, pp.~34--50, 2020.

\bibitem{spwm-introduction}
E.~C. dos.~Santos. Jr and E.~R.~C. da~Silva, ``Optimized PWM approach", in {\em Advanced Power Electronics Converters}. Hoboken, NJ, USA: Wiley, 2014, ch.~8, pp.~221--263.

\bibitem{yokoyama2009}
T.~Yokoyama and T.~Komiyama, ``High-speed single-phase PLL control by quasi dq transformation using FPGA,'' {\em IEEJ Trans. IA}, vol.~129, no.~10, pp.~964--971, 2009.

\bibitem{MIL-STD-704F}
{\em Aircraft Electric Power Characteristics}, MIL-STD-704F, 2004.

\bibitem{blaabjerg2006}
F.~Blaabjerg, R.~Teodorescu, M.~Liserre, and A.~Timbus, ``Overview of control and grid synchronization for distributed power generation systems,'' {\em IEEE Trans. Ind. Electron.}, vol.~53, no.~5, pp.~1398--1409, 2006.

\bibitem{ORTEGA2002585}
R.~Ortega, A.~{van der Schaft}, B.~Maschke, and G.~Escobar, ``Interconnection and damping assignment passivity-based control of port-controlled Hamiltonian systems,'' {\em Automatica}, vol.~38, no.~4, pp.~585--596, 2002.

\bibitem{PRcontroller}
D.~Zmood and D.~Holmes, ``Stationary frame current regulation of PWM inverters with zero steady-state error,'' {\em IEEE Trans. Power Electron.}, vol.~18, no.~3, pp.~814--822, 2003.

\bibitem{outerPItuning}
C.~Bajracharya, M.~Molinas, J.~Suul, and T.~Undeland, ``Understanding of tuning techniques of converter controllers for VSC-HVDC,'' in {\em Proc. Nordic Workshop on Power and Ind. Electron. (NORPIE/2008)}, 2008.

\bibitem{PRtuning}
D.~G. Holmes, T.~A. Lipo, B.~P. McGrath, and W.~Y. Kong, ``Optimized design of stationary frame three phase AC current regulators,'' {\em IEEE Trans. Power Electron.}, vol.~24, no.~11, pp.~2417--2426, 2009.

\end{thebibliography}
\end{document}